\documentclass[twocolumn]{aastex62}

\pdfoutput=1 
\usepackage{amsmath}
\usepackage[T1]{fontenc}

\newcommand{\Ciii}{\ion{C}{3}}
\newcommand{\Civ}{\ion{C}{4}}
\newcommand{\Hii}{\ion{H}{2}}

\newcommand{\Siiii}{\ion{Si}{3}}
\newcommand{\Siiv}{\ion{Si}{4}}
\newcommand{\Lya}{Ly$\alpha$}
\newcommand{\kms}{\ifmmode\,{\rm km}\,{\rm s}^{-1}\else km$\,$s$^{-1}$\fi}

\shorttitle{}
\accepted{1 August 2025}

\begin{document}

\title{Resolved Stellar and Nebular Kinematics of a Star-forming Galaxy at $z \sim 2$}

\author[0009-0007-0184-8176]{Sunny Rhoades}
\affiliation{Department of Physics and Astronomy, University of California, Davis, 1 Shields Avenue, Davis, CA 95616, USA}

\author[0000-0001-5860-3419]{Tucker Jones}
\affiliation{Department of Physics and Astronomy, University of California, Davis, 1 Shields Avenue, Davis, CA 95616, USA}

\author[0000-0002-2645-679X]{Keerthi Vasan G.C.}
\affiliation{The Observatories of the Carnegie Institution for Science, 813 Santa Barbara Street, Pasadena, CA 91101, USA}

\author{Yuguang Chen}
\affiliation{Department of Physics and Astronomy, University of California, Davis, 1 Shields Avenue, Davis, CA 95616, USA}
\affiliation{Department of Physics, The Chinese University of Hong Kong, Shatin, N.T., Hong Kong SAR, China}

\author[0000-0003-4570-3159]{Nicha Leethochawalit}
\affiliation{National Astronomical Research Institute of Thailand (NARIT), Mae Rim, Chiang Mai, 50180, Thailand}

\author{Richard Ellis}
\affiliation{Department of Physics and Astronomy, University College London, Gower Street, London WC1E 6BT, UK}

\author[0000-0002-5558-888X]{Anowar J. Shajib}
\affiliation{Department of Astronomy and Astrophysics, University of Chicago, Chicago, IL 60637, USA}
\affiliation{Kavli Institute for Cosmological Physics, University of Chicago, Chicago, IL 60637, USA}
\affiliation{Center for Astronomy, Space Science and Astrophysics, Independent University, Bangladesh, Dhaka 1229, Bangladesh}

\author[0000-0002-3254-9044]{Karl Glazebrook}
\affiliation{Centre for Astrophysics and Supercomputing, Swinburne University of Technology, PO Box 218, Hawthorn, VIC 3122, Australia}

\author[0000-0001-9676-5005]{Kris Mortensen}
\affiliation{Vera C. Rubin Observatory Project Office, 950 N. Cherry Ave., Tucson, AZ 85719, USA}

\author[0000-0003-4792-9119]{Ryan L. Sanders}
\affiliation{Department of Physics and Astronomy, University of Kentucky, 505 Rose Street, Lexington, KY 40506, USA}

\begin{abstract}
The kinematics of star-forming galaxy populations at high redshifts are integral to our understanding of disk properties, merger rates, and other defining characteristics. Nebular gas emission is a common tracer of galaxies' gravitational potential and angular momenta, but is sensitive to non-gravitational forces as well as galactic outflows, and thus might not accurately trace the host galaxy dynamics. We present kinematic maps of young stars from rest-ultraviolet photospheric absorption in the star-forming galaxy CASSOWARY 13 (a.k.a. SDSS J1237+5533) at $z=1.87$ using the Keck Cosmic Web Imager, alongside nebular emission measurements from the same observations. Gravitational lensing magnification of the galaxy enables good spatial sampling of multiple independent lensed images. We find close agreement between the stellar and nebular velocity fields. We measure a mean local velocity dispersion of $\sigma = 64\pm12$~\kms\ for the young stars, consistent with that of the \Hii\ regions traced by nebular \Ciii] emission ($52\pm9$~\kms). The $\sim 20$~\kms\ average difference in line-of-sight velocity is much smaller than the local velocity width and the velocity gradient ($\gtrsim 100$~\kms). We find no evidence of asymmetric drift nor evidence that outflows bias the nebular kinematics, and thus conclude that nebular emission appears to be a reasonable dynamical tracer of young stars in the galaxy. These results support the picture of star formation in thick disks with high velocity dispersion at $z\sim2$, and represent an important step towards establishing robust kinematics of early galaxies using collisionless tracers. 
\end{abstract}


\section{Introduction}

When and how the modern-day Hubble sequence of morphologies emerged remains an open question in the field of galaxy formation. While passive spheroids and star-forming disks have been observed at high redshifts \citep[e.g.,][]{newman10, wuyts12, ferreira23, jacobs23, kuhn24}, galaxy morphologies at $z\gtrsim2$ are typically irregular and lack the thin-disk spiral structure seen in the local universe \citep[e.g.,][]{elmegreen07, shapley11, guo15, huertas15}.

Modern spiral galaxies such as the Milky Way typically contain a thin disk of young stars, \Hii\ regions, and cold molecular gas \citep[e.g.,][]{yoachim06}, while older stars inhabit a thicker disk \citep[e.g.,][]{bland-hawthorn16}. 
Thin and thick disks are differentiable using both their scale heights and kinematics, with thin disks having smaller velocity dispersion (i.e., kinematically colder). Consequently thin-disk stars have higher measured rotation velocities than thicker-disk stars, a phenomenon referred to as asymmetric drift. Early studies proposed that old stars had originally formed in a thin disk, and subsequently were dynamically heated to comprise modern thick disks \citep{lacey84}. This dynamical heating can be caused by gravitational interactions with dense regions of the disk such as giant star-forming clumps in early disk galaxies \citep{inoue14}, spiral arms, and giant molecular clouds (GMCs). Repeated scattering interactions drive higher velocity dispersions in stars over time, shifting stars originally on thin-disk orbits to a thicker disk on timescales of hundreds of Myr \citep{pessa23}.

However, an alternative scenario is that thick disks form \textit{in situ} at high redshifts \citep{comeron11, hamilton23, tsukui24,hung19}. This picture is supported by kinematics of ionized gas emission lines such as H$\alpha$ which trace star-forming \Hii\ regions \citep[e.g.,][]{ubler24}. Ionized gas velocity dispersions at redshifts $z>1.5$ are typically several times higher than those measured in nearby disk galaxies of comparable mass, suggesting thick-disk kinematics \citep[e.g.,][]{genzel08,jones10a,glazebrook13,wisnioski15,turner17, johnson18,ubler19, wisnioski19}, with the ratio of rotational velocity to dispersion ($V/\sigma$) decreasing over time. This has led to a picture of \space ``disk settling'' whereby disks become thinner at later times, maturing into the modern Hubble sequence at $z\lesssim1$ \citep[e.g.,][]{kassin12,simons17}.

These two models for the origin of thick stellar disks can be summarized as follows. In the first model, the thin disk forms early, and dynamical heating gradually increases the scale height of stellar populations, leading to a thick disk of old stars. In the second, turbulent gas-rich thick disks form \textit{in situ} in galaxies at high redshift, with velocity dispersion decreasing over time leading to progressively thinner star-forming disks at later times \citep[e.g.,][]{grand20, yu23}. This second picture can be understood as a consequence of high gas fractions at early times \citep[e.g.,][]{tacconi20,scoville16}, with galaxies maintaining a marginally stable thick gas disk \citep{elmegreen15}, and delayed formation of a dynamically colder thin disk \citep{vandonkelaar22}. Increased velocity dispersions in ionized gas in high-redshift galaxies also support this latter picture. In contrast, using cold molecular gas as a kinematic tracer supports the former scenario, in which galaxy disks are ``born thin'' and then progressively thicken (e.g., \citealt{rizzo23} find that cold gas velocity dispersions in high-redshift galaxies observed with ALMA can be consistent with modern thin disks). The nature and origin of present-day thick disks thus remains unclear.

Distinguishing these models requires directly establishing the thickness of young stellar disks in the early universe, through scale heights \citep[e.g.,][]{tsukui24} or kinematics. A possible concern with gas kinematics is that measurements of ionized gas velocity dispersion may be biased to higher values by ubiquitous outflows in star-forming galaxies at $z\gtrsim2$ \cite[e.g.,][]{shapley03}, possibly explaining the larger dispersions compared to cold molecular gas \citep[e.g.,][]{ubler19}. Stellar kinematics are robust to effects of outflows, but measurements of the young stars are challenging and limited by low continuum signal-to-noise relative to emission lines from \Hii\ regions.

With integral field unit (IFU) data of gravitationally lensed star-forming galaxies, measuring the kinematics of young stars at high redshift becomes possible. Studies at high redshifts are typically limited by poor spatial resolution, making both morphological and kinematic measurements of star-forming galaxies difficult. Observations of strongly lensed galaxies enable improved spatial sampling of kinematics for galaxies at high redshifts \citep[e.g.,][]{stark08,jones10a,swinbank11,dessauges-zavadsky23}, including stellar kinematics in quiescent galaxies \cite[e.g.,][]{newman18, deugenio24}. Resolved stellar kinematics of massive quiescent galaxies suggest that their stars predominantly formed in a disk \cite[e.g.,][]{toft17, bezanson18}, but the kinematics of young stars in star-forming galaxies are needed to distinguish between thick vs. thin disk formation scenarios.
 
In this paper we present kinematics of young stars in a strongly lensed $z=1.87$ star forming galaxy together with its ionized gas, building on previous studies measuring ionized gas kinematics at similar redshifts. This represents the first application of young-star kinematics to directly investigate the thickness of early stellar disks, and we verify ionized gas as a useful dynamic tracer. The organization of this paper is as follows. In Section~\ref{data}, we discuss the observations and data reduction. Section~\ref{methods} discusses the resolved spectroscopic data and extraction of kinematic properties. In Section~\ref{results} we present 2D kinematic maps, and compare the kinematic properties of young stars and nebular gas, including their local velocity dispersion. We describe the main conclusions in Section~\ref{conclusions}.

\section{Keck/KCWI observations}\label{data}

\begin{figure*}
\centering
    \begin{minipage}{0.9\linewidth}
        \centering
       \includegraphics[width=\linewidth]{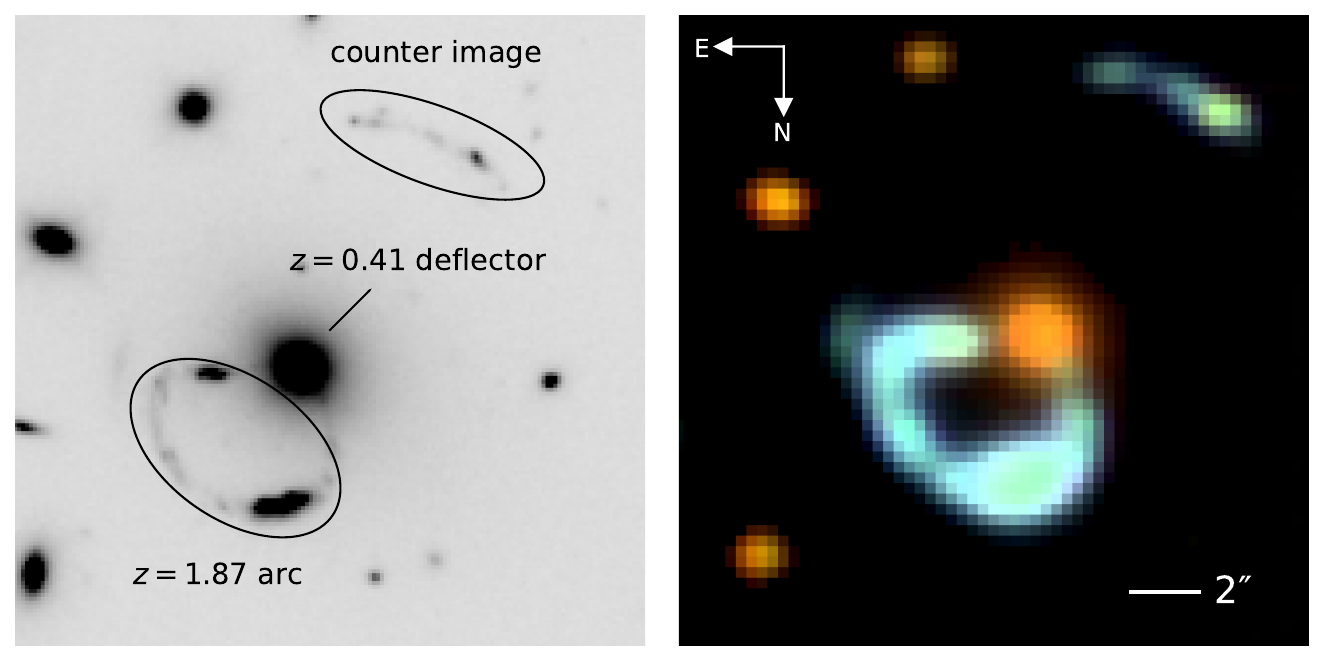}
    \end{minipage}    
    \caption{\emph{Left:} HST/WFC3 image of CSWA 13 in the near-infrared F140W filter. Ovals mark the main arc as well as the counter image. The main deflector is labeled at the center of the frame. \emph{Right:} KCWI color image, showing the prominent blue color of the arc and counter image at optical (rest-frame UV) wavelengths. This image was created following the methods described in \citet{vasan24}, with blue, green, and red color channels obtained by averaging KCWI data over three wavelength ranges (\textit{blue:} 3530--3930~\AA, \textit{green:} 4290--4630~\AA, \textit{red:} 4930--5530~\AA). In both panels North is down and East is to the left.}
    \label{fig:color_img}
\end{figure*}

Our target for this study is CSWA 13, a gravitationally lensed galaxy at $z=1.87$ with stellar mass $\mathrm{log}_{10}(M_*/M_{\odot}) = 9.00 \pm 0.32$ and a star formation rate of $\mathrm{log}_{10}(\mathrm{SFR}/(M_{\odot}\ \mathrm{yr}^{-1})) = 1.71 \pm 0.21$ \citep{vasan24}.
We observed CSWA 13 with the Keck Cosmic Web Imager \citep[KCWI;][]{morrissey18} on the Keck II telescope on 6 April 2022, as described in \cite{vasan24}. Here we provide a summary of the observations and data reduction. Conditions ranged from clear to thin cirrus with seeing of $\simeq$ 1\farcs0. 
Data were taken with the medium slicer using the standard $2 \times 2$ binning, which provides a 16\arcsec\ $\times$ 20\farcs4 field of view sampled with 0\farcs3 $\times$ 0\farcs7 spaxels. We used the BL grating with a central wavelength of 4500~\AA, covering roughly 3500 -- 5700 {\AA} with a spectral resolution of 2.4~\AA\ FWHM ($R\simeq 1900$). This wavelength range includes multiple stellar and nebular spectral features of interest at the redshift of CSWA 13. 
We obtained a total of two hours of integration time, with six 20-minute exposures divided evenly between position angles of 0 and 90 degrees. The two orthogonal position angles allow good sampling of the point-spread function. 
Additionally, we observed a blank-sky region offset roughly 40\arcsec\ from CSWA 13 to use as a background sky subtraction reference. 
Figure \ref{fig:color_img} shows a color image constructed from the KCWI data alongside a Hubble Space Telescope WFC3-IR image \citep{huang25}, illustrating the morphology of the lensed images.
These observations fully cover the bright arc and counter image.

Data were reduced with the IDL-based KCWI data-reduction pipeline\footnote[1]{github.com/Keck-DataReductionPipelines/KcwiDRP}, described in \citet{morrissey18}. In brief, the pipeline subtracts mean dark and bias frames from the raw science images, performs flat-fielding as well as cosmic ray removal, and uses twilight flats to correct for slice-to-slice variance. To produce wavelength-calibrated data cubes from the raw images, the pipeline computes geometry and wavelength solutions using continuum bars and arc lamp calibration frames. The sky background is modeled using our sky offset images. We flux-calibrate our observations of CSWA 13 using the pipeline’s interactive tool and our observed spectra of the standard star BD26D2606 observed taken on June 20, 2020 using the same KCWI configuration. The native rectangular-pixel reduced frames are resampled onto a square-pixel spatial grid of uniform wavelength sampling, corresponding to pixels of dimension 0\farcs3 $\times$ 0\farcs3 $\times$ 1~\AA. 


\begin{figure*}[t]
\centering
    \begin{minipage}[m]{1\linewidth}
        \centering
       \includegraphics[width=\linewidth]{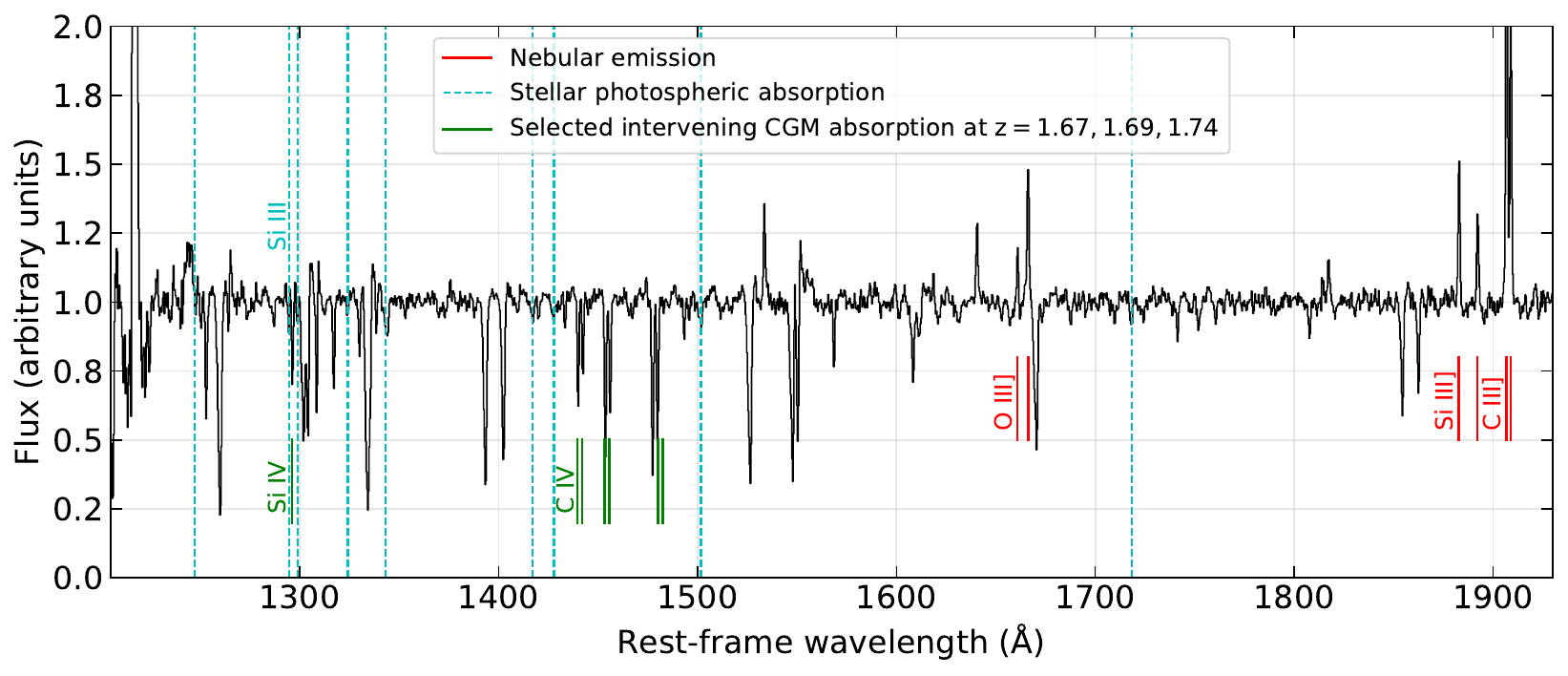}
    \end{minipage}  
    \caption{The integrated, continuum-normalized KCWI spectrum of CSWA 13 in its rest frame ($z=1.86593$), revealing a wealth of features detected with good signal-to-noise. The more prominent stellar photospheric features are marked in cyan, and nebular emission lines in red. Many intervening metal absorption lines are detected, and a subset of the most prominent such features are marked in green. Specifically, we mark the \ion{C}{4} doublets from three distinct absorption systems, and the \ion{Si}{4} $\lambda$1393 absorption line associated with the $z = 1.67$ system. The latter line is blended with \ion{Si}{3} stellar absorption in CSWA 13 at rest-frame 1296~\AA. We additionally detect \Lya\ emission and many strong CGM absorption lines from CSWA 13, and refer readers to \cite{vasan24} for further discussion of these features.}
    \label{fig:1D}
\end{figure*}

\section{Analysis}\label{methods}

Our goal in this work is to establish the stellar velocity structure and compare with that of the \Hii\ regions traced by nebular emission. In this section we determine the systemic redshift and describe the methodology of spatially mapping the velocity fields from stellar and nebular spectroscopic features in CSWA 13. Throughout this analysis we employ the non-linear least squares fitting function \verb|curve_fit| in the Python package \verb|SciPy| to model emission and absorption lines.

\subsection{Systemic redshift from the integrated spectrum}\label{sec:z_sys}

The integrated spectrum of CSWA 13 is shown in Figure~\ref{fig:1D}, created by summing the emission from spaxels corresponding to the bright arc. The spectrum is normalized for display purposes by dividing by a quasi-continuum, constructed by smoothing the observed spectrum with a median filter of 50~\AA. Figure~\ref{fig:1D} shows the observed wavelength range 3500--5500~\AA\ which contains all features of interest for this work, and where KCWI delivers high throughput. We detect several stellar photospheric absorption and nebular emission features as labeled in Figure~\ref{fig:1D}. We also identify multiple strong intervening metal absorption systems (e.g., prominent \Civ\ absorption doublets at $z\simeq1.67$, 1.69, and 1.74), which can result in blending with the intrinsic features of CSWA 13. In this work we avoid such blends. The most relevant intervening absorption in our analysis is the system at $z\simeq1.67$ for which \Siiv~$\lambda$1393 coincides with a region of prominent stellar lines at the redshift of CSWA 13.

To measure the redshift and velocity structure of young stars, we utilize the strongest unblended features available in the KCWI spectrum. 
While several stellar features are detected, many are blends of multiple lines. Examples of prominent stellar features are shown in Figure~\ref{fig:bright_spax}. We identify \Siiii~$\lambda$1294 as the most reliable single-line feature, together with the nearby doublet \Siiii~$\lambda\lambda$1298.89, 1298.96~\AA\ whose two component lines are separated by only 0.07~\AA\ (equivalent to $\simeq$ 15~\kms). We do not use the stellar \ion{Si}{3} feature at 1296~\AA\ for redshift or velocity purposes due to blending with the strong intervening absorption at $z=1.67$. We fit this complex using Gaussian components for the $\lambda$1294 and $\lambda\lambda$1298 stellar features which are assumed to have the same redshift and line width, together with a linear continuum and an independent Gaussian to capture the blend of intervening and stellar absorption at rest-frame 1296~\AA. The strong interstellar \ion{O}{1}~$\lambda$1302 and \ion{Si}{2}~$\lambda$1304 are simultaneously fit with Gaussian components, which allows the full region around \Siiii~$\lambda\lambda$1298.89, 1298.96 to be included, thus improving the results. Altogether this amounts to fitting 5 absorption components plus a linear continuum, where the key stellar features at 1294 and 1298~\AA\ are jointly constrained. 
We find that this provides an adequate fit to the stellar absorption. The best-fit stellar systemic redshift along with the intervening absorption redshift are given in Table~\ref{table:1}. 
Additionally in Table~\ref{table:1} we report redshifts measured from Gaussian fits to the blended stellar photospheric features at 1323 and 1718~\AA, shown in Figure~\ref{fig:bright_spax}. 

The systemic redshift can also be measured from various nebular emission lines detected at high significance. The \Ciii]~$\lambda\lambda$1907, 1909 doublet is the strongest of these (Figure~\ref{fig:1D}). We fit \Ciii] with a double Gaussian profile assuming the same redshift and width for both lines and report the best-fit redshift in Table~\ref{table:1}. The stellar and nebular redshifts are consistent within the $1\sigma$ measurement uncertainty, corresponding to a difference of $14 \pm 25$~\kms. 
As noted above, we consider the \ion{Si}{3} 1294 and 1298~\AA\ features to be the most reliable tracers of stellar kinematics since the others are blends of unresolved lines.
We note that these are flux-weighted average redshifts of the galaxy, whereas a spatial average (i.e., taking the mean redshift of all spaxels; Section~\ref{sec:kinematics}) gives a similar difference of 24~\kms. 
The stellar and nebular redshifts may be expected to differ slightly if the flux distribution of nebular emission and stellar continuum are not the same (e.g., if \Ciii] emission has higher equivalent width on one side of the galaxy), given the observed velocity gradient described in Section~\ref{sec:kinematics}.

\begin{deluxetable}{p{2.7cm}|p{2.5cm}|p{2.5cm}}
    \tablecaption{Redshifts for selected features measured in the integrated spectrum of CSWA 13.} 
    \tablewidth{0.45\textwidth}
    \tabletypesize{\small}
    \tablehead{\colhead{Transition}  & \colhead{Origin} & \colhead{$z_{\mathrm{sys}}$}}
    \startdata 
    \ion{Si}{3} 1294, 1298 \AA & stellar absorption & $1.86593 \pm  0.00024$ \\
    \ion{N}{3}, \ion{C}{2} 1323, 1324 \AA & stellar absorption & $1.86542 \pm 0.00023$ \\
    \ion{Fe}{4}, \ion{N}{4} $ 1718$ \AA & stellar absorption & $1.86504 \pm 0.00014$ \\
    \Ciii] 1907, 1909 \AA & nebular emission & $1.86606 \pm 0.00002$ \\
    \Siiii] 1882, 1892 \AA & nebular emission & $1.86613 \pm 0.00003$ \\
    \ion{Si}{4} $1393$\AA & intervening absorption$^a$ & $1.66585 \pm 0.00016$
    \enddata
    \tablenotetext{}{$^a$We detect spectral lines from several intervening absorbers at $z = 1.67$, 1.69, and 1.74, as also described in \citet{vasan24}. The \ion{Si}{4} feature at $z = 1.67$ is blended with the \ion{Si}{3} stellar line at 1296 \AA\ (see Figure~\ref{fig:bright_spax}); here we report the redshift corresponding to \ion{Si}{4}~$\lambda$1393.}
\vspace{-20pt} 
\label{table:1}
\end{deluxetable}


\subsection{Kinematic properties}
\label{sec:kinematics}

We now turn to the spatially resolved velocity field of CSWA 13. To improve the signal-to-noise ratio of stellar absorption lines, we first lightly smooth the KCWI data cube with a 2D Gaussian kernel with FWHM of 0\farcs9 (3 spaxels). This results in an effective angular resolution of approximately 1\farcs3 FWHM. We use an object mask to select spaxels in the main arc with sufficient continuum flux density to allow reliable detection of stellar features. 
The error spectrum is determined at each wavelength pixel from the standard deviation of a blank region of sky to the northeast of the main arc, where no objects are detected in the KCWI data cube. This error spectrum is consistent with the noise level of individual spaxels in relatively featureless regions of the CSWA 13 spectrum, and we adopt it for purposes of curve fitting and quantifying uncertainties. 

\begin{figure}[h]
    \centering
    \includegraphics[width=.47\textwidth]{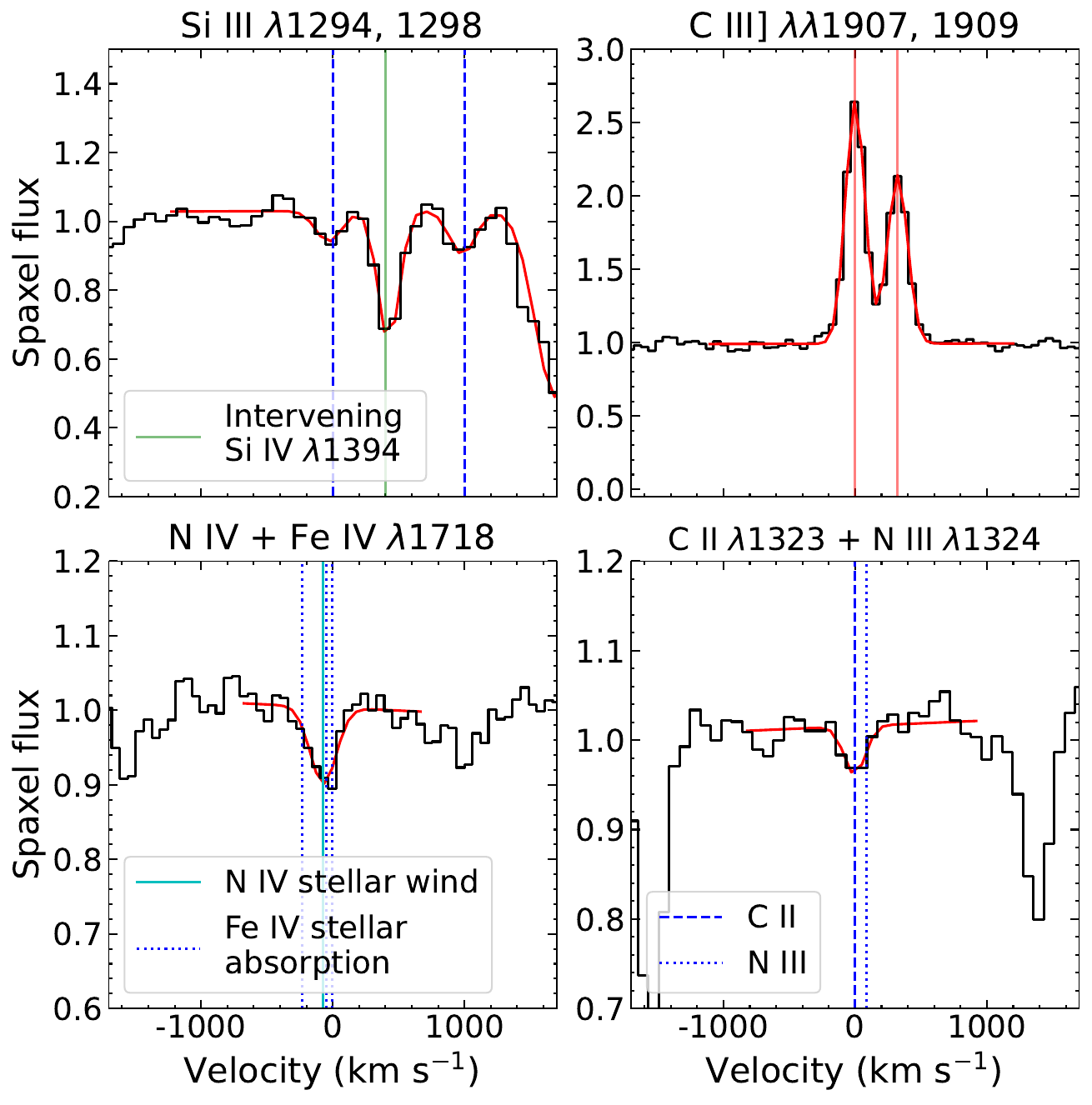}
    \caption{The spectrum in a single bright spaxel of CSWA 13, zoomed in around stellar and nebular features of interest: stellar \Siiii\ along with intervening absorption from \Siiv\ at $z=1.67$ (\textit{top left}), \Ciii] nebular emission (\textit{top right}), blended \ion{N}{4} and \ion{Fe}{4} stellar absorption (\textit{bottom left}), and blended \ion{N}{3} with \ion{C}{2} stellar absorption (\textit{bottom right}). In each panel the spectrum is shown in black with best-fit profiles in red. Vertical lines mark the location of each spectral line, with legends distinguishing the different species of blended features. The velocity origin corresponds to an individual line in each panel.}
    
    \label{fig:bright_spax}
\end{figure}

For each selected spaxel in the main arc, we fit the stellar \ion{Si}{3} absorption lines at 1294 and 1298~\AA\ together with the nearby intervening and interstellar features, as described in Section~\ref{sec:z_sys}. We impose bounds on the redshift and line width in order to avoid spurious fits. 
This provides spatial maps of the stellar velocity ($V_\mathrm{Si~ \sc{III}}$) and velocity dispersion ($\sigma_\mathrm{Si~ \sc{III}}$) in each spaxel. (We note that the fit also provides maps of interstellar gas kinematics, discussed in detail by \citealt{vasan24}.) An example fit to a single bright spaxel is shown in Figure~\ref{fig:bright_spax}. We also show single-component fits to the stellar features at 1718 and 1323~\AA.
We likewise fit nebular \Ciii] emission following the same method as for the integrated spectrum, yielding maps of the nebular velocity and dispersion. \Ciii] has far higher signal-to-noise than the stellar features (as can be seen in Figures~\ref{fig:1D} and \ref{fig:bright_spax}), and we are thus able to measure the nebular kinematics over a slightly larger area.

\begin{figure*}
\centering
    \begin{minipage}{1\linewidth}
        \centering
       \includegraphics[width=\linewidth]{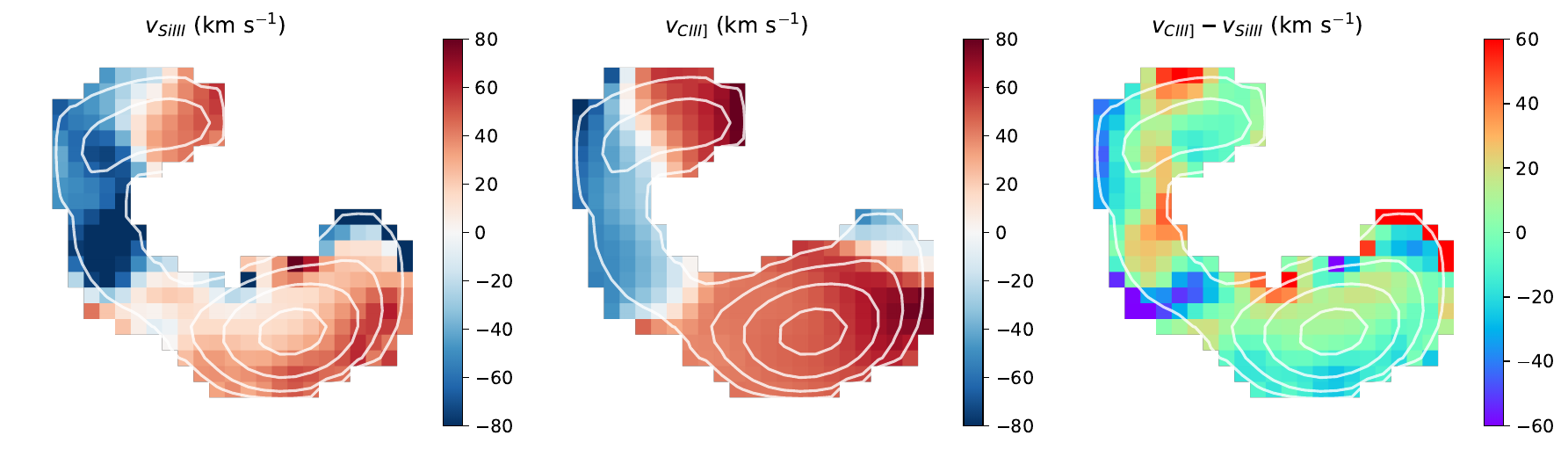}
    \end{minipage}   
    \caption{Maps of the velocity field measured in the image plane for stellar \ion{Si}{3} absorption (\textit{left}) and nebular \Ciii] emission (\textit{center}), along with their difference (\textit{right}, with 24~\kms\ offset removed). White contours show continuum flux density levels measured from the same KCWI data. The nebular and stellar velocity maps show similar overall structure, with a blue-to-redshifted gradient seen mirrored in multiple lensed images (see \citealt{vasan24} for discussion of the intrinsic source-plane structure). The stellar velocity is offset by a $24$~\kms \space average, with a scatter of $31$~\kms.}
    \label{fig:panel}
\end{figure*}

Figure \ref{fig:panel} shows the velocity maps of young stars (traced by \Siiii\ photospheric absorption) and nebular gas (traced by \Ciii] emission) in CSWA 13, as seen in the lensed image plane for the main arc. 
The stellar and nebular maps exhibit similar velocity gradients of $>100$~\kms\ across the three multiple images of the lensed galaxy. The de-lensed image shows a smooth velocity profile in the source plane, compatible with a rotating disk (Figure \ref{fig:rot_curve}). We refer readers to \citet{vasan24} for a detailed description and figures of the source-plane structure (e.g., Region A is redshifted and Region C is blueshifted). Here we focus on the stellar kinematics and comparison between stellar and nebular measurements. We adopt the image plane for our analysis since the point-spread function and noise properties are easily characterized. The main conclusions of this paper hold for both the de-lensed source plane and the observed image plane. 

\section{Results and discussion}\label{results}

In this section, we characterize the stellar and nebular velocity structure across the star-forming galaxy CSWA 13 and present spatially-resolved kinematics maps of each tracer. Spatial maps of the velocity fields of the young stars and the ionized gas are similar in structure, as shown in three independent images of the lensed galaxy. Comparing stellar and nebular rotational velocities in 1D, we find these two tracers to be consistent within their respective measurement uncertainties. 
We measure a consistent velocity dispersion for the young stars and nebular gas in CSWA 13, which is typical of star-forming galaxies at this redshift.

\subsection{Velocity structure}\label{v-structure}

CSWA 13 shows a well-resolved velocity structure. A visual inspection of both stellar and nebular velocity fields reveals a clear velocity gradient suggesting disk rotation. The left and center panels of Figure~\ref{fig:panel} show the corresponding 2D maps of stellar and nebular velocity fields in the image plane around three multiply-lensed regions of the arc (Figure~\ref{fig:color_img}). 

Despite overall structural agreement in our comparison of the stellar and nebular gas velocity fields, we find a slight difference between the \ion{Si}{3} and \Ciii] observed velocities (see right panel of Figure~\ref{fig:panel}). On average, velocity measurements of \Ciii] emission are redshifted relative to stellar absorption by 24~\kms. Although \ion{Si}{3} absorption is not as well-detected across the arc as \Ciii] emission, we do not find that spurious fits to the \ion{Si}{3} complex affect our comparison, as the brightest regions show the most statistically significant difference, $2.7\sigma$ between measured nebular and stellar velocities. If outflows are contributing significantly to the \Ciii] kinematics, we would expect a blueshifted profile due to greater attenuation of the redshifted outflow component \citep[e.g.,][]{soto2012} in contrast to the observed average redshift. Thus we conclude that this average difference is likely not caused by outflowing gas.

\begin{figure}[h]
    \centering
    \includegraphics[width=\linewidth]{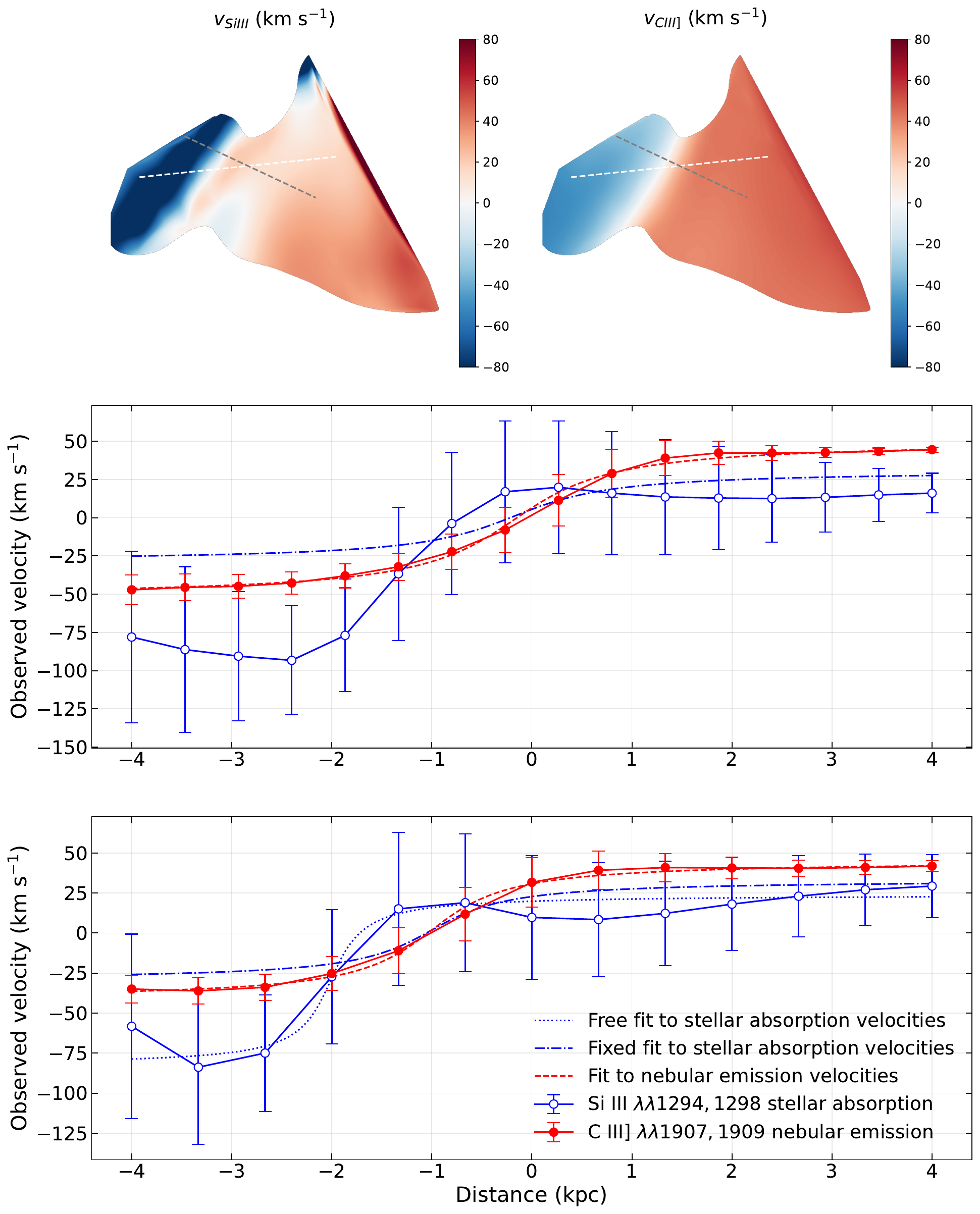}
    \caption{\emph{Top:} Source-plane velocity maps for the highly magnified central region in the CSWA 13 arc, reconstructed from stellar (\textit{left}) and nebular (\textit{right}) line measurements. \emph{Middle:} Position-velocity diagram for nebular emission and photospheric absorption, extracted along the pseudo-slits shown as dashed white lines in the upper panels. 
    This pseudo-slit position is chosen to sample the brightest regions, providing high signal-to-noise near the kinematic major axis.
    Dashed blue and red lines show arctangent fits to the stellar and nebular velocity curves. 
    The stellar fit adopts the same center, systemic velocity, and scale length as the nebular emission profile which has higher signal-to-noise ratio. \emph{Bottom:} Position-velocity diagram extracted along the dashed gray pseudo-slit shown in the upper panels, which matches the approximate orientation of the kinematic major axis. We use the same color and line conventions for the nebular emission and stellar absorption velocity curves and resulting arctangent fits as in the middle panel, additionally including a free fit to the stellar velocity curve that is not fixed to nebular emission profile parameters (\textit{dotted blue line}). While we observe an average blueshift of stellar lines relative to nebular emission for both choices of pseudo-slit orientation, the best-fit asymptotic velocities for the nebular gas and young stars remain consistent. 
    }
    \label{fig:rot_curve}
\end{figure}

In Figure~\ref{fig:rot_curve}, we construct source-plane position-velocity diagrams for the central image of CSWA 13 showing both stellar and nebular gas kinematics, using the same lens model and methodology as in \citet{vasan24}. We extract velocities along the major axis pseudo-slits shown in both spatial maps. The pseudo-slit matched to the orientation of the kinematic major axis probes fainter, lower signal-to-noise regions of the arc compared those sampled by the oblique slit. We use the oblique slit in our analyses as a consequence of this, but find measurements from both slits to be consistent within $1\sigma$ uncertainties.
The average redshift of nebular gas relative to stars is apparent in the 1-D velocity profiles, although with limited statistical significance for individual spatial resolution elements. Both stars and nebular gas show a similar velocity shear of $\Delta V \sim 100$~\kms, suggesting $V_{rot} \sin{i} \sim 50$~\kms\ for circular rotation velocity $V_{rot}$ with disk inclination angle $i$.

Figure~\ref{fig:rot_curve} allows us to quantify the degree of asymmetric drift (if any) in CSWA 13. In low-redshift disk galaxies, asymmetric drift is commonly observed as a lag between the average rotational velocity of gas and that of the stars \citep[e.g.,][]{blanc13, martinsson13, bershady24}, with slower rotation of the stars explained by their larger velocity dispersion. In a sample of $z\sim 0$ disk galaxies with optical integral-field spectroscopy from the Sloan Digital Sky Survey, \cite{bershady24} measure a median difference $V_a$ between the tangential speed of stars and ionized gas of $16.5$~\kms. 

At high redshift, if the nebular gas velocity dispersion is larger than that of the young stars, we may expect an opposite asymmetric drift effect. We measure the asymmetric drift between young stars and nebular gas within the central image of CSWA 13 using an arctangent function to estimate the asymptotic velocity for both position-velocity curves in Figure~\ref{fig:rot_curve}. 
We first fit the nebular gas rotation curve, and fix the stellar rotation curve to have the same center, systemic velocity, and scale length (the ``fixed fit''). We obtain consistent results when allowing these parameters to be free (the ``free fit''), albeit with larger uncertainty in the best-fit asymptotic velocity of the stellar rotation curve 
($53 \pm 27$~\kms\ for the free fit cf. $30 \pm 9$~\kms\ for the fixed fit). 
The resulting arctangent profile fits are shown in Figure~\ref{fig:rot_curve}. 
We quantify the goodness-of-fit by comparing the RMS difference to the typical error for both the young stars and nebular gas measurements. 
For the stellar fixed fits we measure RMS~$\simeq 30$~\kms\ which is comparable to the typical measurement uncertainty ($\simeq 39$~\kms), indicating a good fit. The stellar free fit has RMS~$=7$~\kms, considerably smaller than the measurement uncertainties, and we likewise find a relatively small RMS for the nebular fits. The small RMS for those fits with many free parameters is likely a result of the modest number of independent seeing resolution elements across the pseudo-slits, such that there are few degrees of freedom.
Subtracting best-fitting asymptotic velocities, we measure a difference of $V_a = -22 \pm 20$~\kms\ between the stars and gas. 
The data are therefore consistent with no asymmetric drift ($V_a = 0$) between the gas and young stars. We note that the 20~\kms\ precision of our $V_a$ measurement is comparable to the asymmetric drift found between nebular gas and older stars in nearby disk galaxies \citep{bershady24}. 
Applying the same analysis to a source-plane reconstruction of the southern main arc image of CSWA 13, we find that the difference in stellar and nebular asymptotic velocities is consistent within uncertainties to that measured for the central image. 
Overall we thus find that the stellar and nebular kinematics are in general agreement, with a modest average offset.


\subsection{Velocity dispersions}\label{v-disp}

\begin{figure}[h]
    \centering
    \includegraphics[width=0.4\textwidth]{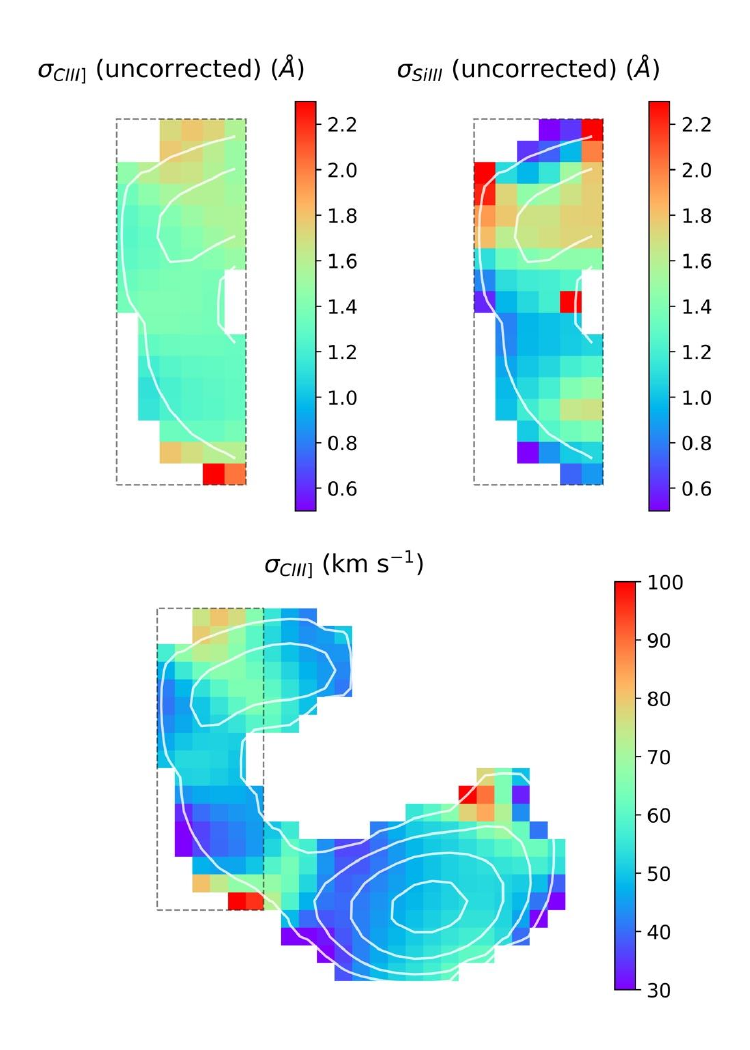}
    \caption{Image-plane spatial maps of the measured line widths (uncorrected for KCWI instrumental broadening) for \Ciii] nebular gas (\textit{top left}) and the stellar \Siiii\ velocity dispersion (\textit{top right}). The instrumental resolution is $\sigma_{inst} = 1.02$~\AA. For reference, a velocity dispersion of $\sigma = 50$~\kms\ corresponds to 1.19~\AA\ for the stellar \Siiii\ lines and 1.37~\AA\ for nebular \Ciii]. We find good agreement on average, with larger uncertainty and noticeably larger scatter for the stellar fits. 
    We note that best-fit \Siiii\ line widths are sometimes smaller than the instrument resolution in spaxels with low continuum signal-to-noise, corresponding to unphysical velocity dispersions. We measure a physical result for \Ciii] in nearly every spaxel thanks to its higher signal-to-noise, and show the resulting intrinsic velocity dispersion map in the \textit{bottom} panel. To account for unphysical best fits to \Siiii, we report an average value of the local stellar velocity dispersion derived from measured line widths within a subsection of the arc. The region of the arc used to estimate the local \Ciii] and \Siiii\ velocity dispersions corresponds to the top panels and is indicated by dashed gray boxes. 
    Continuum contours are overlaid in white. We find a mean velocity dispersion of $\sigma = 52$~\kms\ for nebular \Ciii] emission, and $\sigma = 64$~\kms\ for stellar \Siiii\ absorption. The largest deviations from these averages are in regions of low signal-to-noise near the edge of the arc, and are largely explained by measurement uncertainty.}
    \label{fig:c3_dispersion}
\end{figure}

We now examine the velocity dispersion of nebular gas and young stars in CSWA 13. In Figure \ref{fig:c3_dispersion} we show the 2D maps of \Ciii] and \Siiii\ line widths from which we calculate velocity dispersion, along with the intrinsic velocity dispersion map for nebular \Ciii].
To correct line widths for instrumental broadening, we assume a Gaussian profile with FWHM $= 2.4$~\AA\ as described in \citet{mortensen21}. Fitted line widths below this instrumental resolution are formally non-physical, but are sometimes obtained in spaxels with lower signal-to-noise.
Measured values typically correspond to $\sigma=30$--80~\kms. To estimate the local velocity dispersion $\sigma_0$, we take the average value within a region of the arc with minimal velocity shear in order to minimize beam-smearing effects. Specifically, we use the blueshifted central region of the main arc. This region is representative in terms of velocity dispersion (Figure~\ref{fig:c3_dispersion}) and has the advantage of a larger separation between the stellar \Siiii\ and interstellar \ion{O}{1}~$\lambda$1302 absorption. We measure a mean local velocity dispersion $\sigma_0 = 52\pm9$~\kms\ for the nebular gas and $64\pm12$~\kms\ for the young stars, consistent within the 1$\sigma$ uncertainty. 

\begin{figure}
    \centering
    \includegraphics[width=0.45\textwidth]{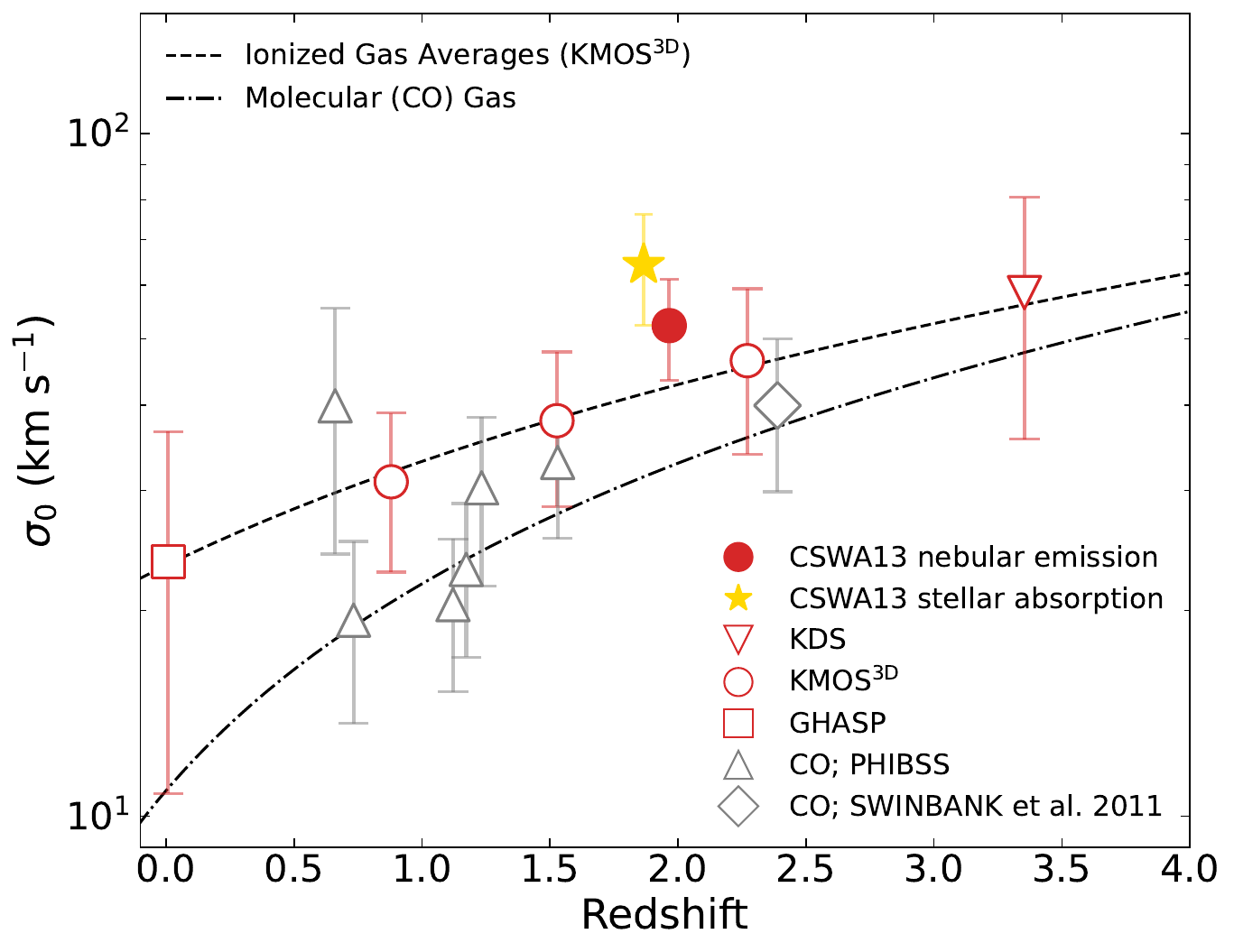}
    \caption{The evolution of intrinsic velocity dispersion $\sigma_0$ with redshift in star-forming galaxies, adapted from Figure 6 in \cite{ubler19}. Open red triangles, circles, and squares show the results from nebular gas surveys using H$\alpha$, including our measurement with \Ciii] in CSWA 13 (solid red circle, off-set in redshift for ease of comparison). Open grey triangles and diamonds show the results from molecular gas studies \citep{tacconi13,freundlich19,swinbank11}. Our stellar velocity dispersion measurement for CSWA 13 is denoted by a gold star. CSWA 13 exhibits a typical nebular gas $\sigma_0$ for its redshift, while the stellar $\sigma_0$ is slightly larger and consistent within the measurement uncertainties. These results support the picture of relatively large velocity dispersion for young stars, indicating thick disk kinematics, as opposed to the smaller dispersions implied by molecular gas measurements.}
    \label{fig:sig0-z-fig}
\end{figure}

We compare the velocity dispersion of CSWA 13 with measurements from the literature in Figure~\ref{fig:sig0-z-fig}. The velocity dispersion from nebular emission is in good agreement with the average trend at $z\sim2$ from the large KMOS$^{\mathrm{3D}}$ survey \citep{wisnioski19,ubler19}. Using results from \citet{mainali23}, \citet{vasan24} derive a magnification-corrected stellar mass and SFR of $\mathrm{log}_{10}(M_*/M_{\odot}) = 9.00 \pm 0.32$ and $\mathrm{log}_{10}(\mathrm{SFR}/(M_{\odot}\ \mathrm{yr}^{-1})) = 1.71 \pm 0.21$, which places CSWA 13 at lower mass (cf. $\mathrm{log}_{10}(M_*/M_{\odot}) \gtrsim 9.5$) and higher sSFR compared to $z\sim1.5-2$ galaxies from the KMOS$^{\mathrm{3D}}$ sample. While CSWA 13 has relatively low stellar mass compared to KMOS$^{\mathrm{3D}}$ galaxies, it is evidently typical in terms of $\sigma_0$. We note that the treatment of beam smearing effects differs between our reported values and those of \cite{ubler19}, such that the measurement methods are similar but not strictly identical. 

A key question for this work is whether the nebular emission kinematics are an accurate tracer of the young stars, particularly for the local velocity dispersion $\sigma_0$. Measurements of cold gas at high redshift show evidence for smaller velocity dispersion than the nebular gas (Figure~\ref{fig:sig0-z-fig}; e.g., \citealt{tacconi13,freundlich19,ubler18,rizzo24}). These cold gas measurements suggest that stars may form in thinner disks, as opposed to the thick disks implied by large nebular velocity dispersions. In this scenario we would expect $\sigma_0$ for the young stars to also be smaller than the nebular gas value, and closer to that of the cold gas. We instead find that $\sigma_0$ measured from young stars is slightly larger (and consistent within 1$\sigma$) with that of the nebular gas. Our young star value $\sigma_0 = 64\pm12$~\kms\ is also larger than the typical $\sim$30~\kms\ for cold molecular gas at $z\sim2$ (Figure~\ref{fig:sig0-z-fig}; \citealt{ubler19}). Our measurements therefore do not support the picture of stars forming in relatively thin disks in typical $z\sim2$ galaxies, unless there is significant dynamical thickening over the $\sim$100~Myr lifetimes of the stars traced by the rest-UV \Siiii\ features \citep[e.g.,][]{hamiltoncampos23,pessa23}. Instead we find direct evidence of high velocity dispersion indicative of a thick disk for the young stellar population in CSWA 13, consistent with the nebular gas kinematics.

\section{Conclusions}\label{conclusions}

In this paper, we present spatially resolved kinematics of young stars and nebular gas (via \Ciii] emission) in CSWA 13, a strongly star-forming $\mathrm{log}_{10}(M_*/M_{\odot}) \simeq 9$ galaxy at $z=1.87$ which is lensed by a foreground group. We demonstrate the utility of sensitive rest-UV integral field spectroscopy to measure stellar velocity structure, with the aid of gravitational lensing. This study benefits from measuring the key stellar and nebular features in the same data set from Keck/KCWI. This allows us to directly assess the robustness of ionized gas as a dynamical tracer relative to the stars. Our main results are as follows.

\begin{itemize}
  \item Comparing the velocity fields of the young stars and nebular gas, we observe similar bulk motion in multiple images of the arc suggestive of a rotating disk (e.g., Figure~\ref{fig:rot_curve}). The line-of-sight velocity gradient is $\gtrsim 100$~\kms\ peak-to-peak, and in spatial maps we find average differences of only $\sim$20~\kms\ between the stellar and nebular velocity fields.
  \item We measure the local velocity dispersion from regions of the galaxy with minimal velocity shear, and find consistent values for the young stars ($\sigma_{0,\mathrm{stars}} = 64\pm12$~\kms) and nebular gas ($\sigma_{0,\mathrm{nebular}} = 52\pm9$~\kms). We see no evidence of outflows affecting measurements of nebular gas velocity dispersions in CSWA 13, which would be expected to cause higher $\sigma_0$ for the gas relative to the young stars. We likewise find no evidence for asymmetric drift between the young star and nebular velocity fields, as expected if they have similar velocity dispersions. Our measurement of asymmetric drift in CSWA 13 ($V_a = -22 \pm 20$~\kms) has precision comparable to typical values seen in older stellar populations within $z\sim0$ disk galaxies \citep[$V_a \simeq 16$~\kms;][]{bershady24}. This measurement is the first of its kind at $z \sim 2$.
  We suggest that peculiar velocities of individual star-forming regions and young star clusters in the galaxy might explain the moderate differences ($\sim$20~\kms) in the velocity fields, in that OB complexes and \Hii\ regions may have different random motions with both having overall velocity dispersion $\sigma \simeq 60$~\kms.
  \item We compare the nebular and stellar velocity dispersions of CSWA 13 with average trends in $\sigma_0$ from several kinematic samples of disk galaxies of comparable mass and redshift (Figure~\ref{fig:sig0-z-fig}). The nebular gas velocity dispersion in CSWA 13 is consistent with averages from larger samples (e.g., KMOS$^{\mathrm{3D}}$; \citealt{ubler19}), whose relatively high $\sigma_0$ values have been interpreted as evidence of thick star-forming disks \citep[e.g.,][]{comeron19, bird21, hamiltoncampos23}. The nebular and stellar velocity dispersions are larger than average measurements of molecular gas at similar redshift, possibly indicating rapid dynamical heating of the young stars. The consistent stellar and nebular velocity dispersion measured for CSWA 13 in this work supports the picture in which stars at high redshift form \textit{in situ} in thick disks.
\end{itemize}

In this work, we have successfully demonstrated that kinematics of young stars are accessible at high redshift especially in gravitationally-lensed galaxies, and can address the question of whether stars form in thin or thick disks. From our study of CSWA 13 at $z=1.87$, we find that the velocity dispersions of young stars and nebular emission are similar and indicative of relatively thick disks, comparable to nebular emission measurements of larger samples at similar redshift. We intend to continue this work at higher spectral resolution to better understand the stellar structure and kinematics of galaxies representing the broader population at moderate to high redshifts.

\section*{Acknowledgements}
We thank the anonymous referee for a constructive review which has improved the content and clarity of this manuscript. 
The data presented herein were obtained at the W. M. Keck Observatory, which is operated as a scientific partnership among the California Institute of Technology, the University of California and the National Aeronautics and Space Administration. The Observatory was made possible by the generous financial support of the W. M. Keck Foundation.
TJ and SR gratefully acknowledge support from the National Science Foundation through grant AST-2108515, the Gordon and Betty Moore Foundation through Grant GBMF8549, and a Dean’s Faculty Fellowship and Chancellor's Fellowship.
KVGC was supported by NASA through the STScI grants JWST-GO-04265 and JWST-GO-03777.
KG acknowledges support from Australian Research Council Centre of Excellence CE170100013 and Discovery Project DP230101775.
RSE acknowledges generous financial support from the Peter and Patricia Gruber Foundation.
AJS received support from NASA through the STScI grants HST-GO-16773 and JWST-GO-2974.
The authors wish to recognize and acknowledge the very significant cultural role and reverence that the summit of Maunakea has always had within the indigenous Hawaiian community. We are most fortunate to have the opportunity to conduct observations from this mountain.

\bibliography{draft}

\begin{thebibliography}{}
\expandafter\ifx\csname natexlab\endcsname\relax\def\natexlab#1{#1}\fi
\providecommand{\url}[1]{\href{#1}{#1}}

\bibitem[{{Bershady} {et~al.}(2024){Bershady}, {Westfall}, {Shetty}, {Law},
  {Cappellari}, {Drory}, {Bundy}, \& {Yan}}]{bershady24}
{Bershady}, M.~A., {Westfall}, K.~B., {Shetty}, S., {et~al.} 2024, \mnras, 531,
  1592

\bibitem[{{Bezanson} {et~al.}(2018){Bezanson}, {van der Wel}, {Pacifici},
  {Noeske}, {Bari{\v{s}}i{\'c}}, {Bell}, {Brammer}, {Calhau}, {Chauke}, {van
  Dokkum}, {Franx}, {Gallazzi}, {van Houdt}, {Labb{\'e}}, {Maseda},
  {Mu{\~n}os-Mateos}, {Muzzin}, {van de Sande}, {Sobral}, {Straatman}, \&
  {Wu}}]{bezanson18}
{Bezanson}, R., {van der Wel}, A., {Pacifici}, C., {et~al.} 2018, \apj, 858, 60

\bibitem[{{Bird} {et~al.}(2021){Bird}, {Loebman}, {Weinberg}, {Brooks},
  {Quinn}, \& {Christensen}}]{bird21}
{Bird}, J.~C., {Loebman}, S.~R., {Weinberg}, D.~H., {et~al.} 2021, \mnras, 503,
  1815

\bibitem[{{Blanc} {et~al.}(2013){Blanc}, {Weinzirl}, {Song}, {Heiderman},
  {Gebhardt}, {Jogee}, {Evans}, {van den Bosch}, {Luo}, {Drory}, {Fabricius},
  {Fisher}, {Hao}, {Kaplan}, {Marinova}, {Vutisalchavakul}, \&
  {Yoachim}}]{blanc13}
{Blanc}, G.~A., {Weinzirl}, T., {Song}, M., {et~al.} 2013, \aj, 145, 138

\bibitem[{{Bland-Hawthorn} \& {Gerhard}(2016)}]{bland-hawthorn16}
{Bland-Hawthorn}, J., \& {Gerhard}, O. 2016, \araa, 54, 529

\bibitem[{{Comer{\'o}n} {et~al.}(2019){Comer{\'o}n}, {Salo}, {Knapen}, \&
  {Peletier}}]{comeron19}
{Comer{\'o}n}, S., {Salo}, H., {Knapen}, J.~H., \& {Peletier}, R.~F. 2019,
  \aap, 623, A89

\bibitem[{{Comer{\'o}n} {et~al.}(2011){Comer{\'o}n}, {Elmegreen}, {Knapen},
  {Salo}, {Laurikainen}, {Laine}, {Athanassoula}, {Bosma}, {Sheth}, {Regan},
  {Hinz}, {Gil de Paz}, {Men{\'e}ndez-Delmestre}, {Mizusawa},
  {Mu{\~n}oz-Mateos}, {Seibert}, {Kim}, {Elmegreen}, {Gadotti}, {Ho},
  {Holwerda}, {Lappalainen}, {Schinnerer}, \& {Skibba}}]{comeron11}
{Comer{\'o}n}, S., {Elmegreen}, B.~G., {Knapen}, J.~H., {et~al.} 2011, \apj,
  741, 28

\bibitem[{{Dessauges-Zavadsky} {et~al.}(2023){Dessauges-Zavadsky}, {Richard},
  {Combes}, {Messa}, {Nagy}, {Mayer}, {Schaerer}, {Egami}, \&
  {Adamo}}]{dessauges-zavadsky23}
{Dessauges-Zavadsky}, M., {Richard}, J., {Combes}, F., {et~al.} 2023, \mnras,
  519, 6222

\bibitem[{{D'Eugenio} {et~al.}(2024){D'Eugenio}, {P{\'e}rez-Gonz{\'a}lez},
  {Maiolino}, {Scholtz}, {Perna}, {Circosta}, {{\"U}bler}, {Arribas},
  {B{\"o}ker}, {Bunker}, {Carniani}, {Charlot}, {Chevallard}, {Cresci},
  {Curtis-Lake}, {Jones}, {Kumari}, {Lamperti}, {Looser}, {Parlanti}, {Rix},
  {Robertson}, {Rodr{\'\i}guez Del Pino}, {Tacchella}, {Venturi}, \&
  {Willott}}]{deugenio24}
{D'Eugenio}, F., {P{\'e}rez-Gonz{\'a}lez}, P.~G., {Maiolino}, R., {et~al.}
  2024, Nature Astronomy, 8, 1443

\bibitem[{Elmegreen \& Hunter(2015)}]{elmegreen15}
Elmegreen, B.~G., \& Hunter, D.~A. 2015, The Astrophysical Journal, 805, 145.
\newblock \url{https://dx.doi.org/10.1088/0004-637X/805/2/145}

\bibitem[{{Elmegreen} {et~al.}(2007){Elmegreen}, {Elmegreen}, {Ravindranath},
  \& {Coe}}]{elmegreen07}
{Elmegreen}, D.~M., {Elmegreen}, B.~G., {Ravindranath}, S., \& {Coe}, D.~A.
  2007, \apj, 658, 763

\bibitem[{{Ferreira} {et~al.}(2023){Ferreira}, {Conselice}, {Sazonova},
  {Ferrari}, {Caruana}, {Tohill}, {Lucatelli}, {Adams}, {Irodotou}, {Marshall},
  {Roper}, {Lovell}, {Verma}, {Austin}, {Trussler}, \& {Wilkins}}]{ferreira23}
{Ferreira}, L., {Conselice}, C.~J., {Sazonova}, E., {et~al.} 2023, \apj, 955,
  94

\bibitem[{{Freundlich} {et~al.}(2019){Freundlich}, {Combes}, {Tacconi},
  {Genzel}, {Garcia-Burillo}, {Neri}, {Contini}, {Bolatto}, {Lilly},
  {Salom{\'e}}, {Bicalho}, {Boissier}, {Boone}, {Bouch{\'e}}, {Bournaud},
  {Burkert}, {Carollo}, {Cooper}, {Cox}, {Feruglio}, {F{\"o}rster Schreiber},
  {Juneau}, {Lippa}, {Lutz}, {Naab}, {Renzini}, {Saintonge}, {Sternberg},
  {Walter}, {Weiner}, {Wei{\ss}}, \& {Wuyts}}]{freundlich19}
{Freundlich}, J., {Combes}, F., {Tacconi}, L.~J., {et~al.} 2019, \aap, 622,
  A105

\bibitem[{{Genzel} {et~al.}(2008){Genzel}, {Burkert}, {Bouch{\'e}}, {Cresci},
  {F{\"o}rster Schreiber}, {Shapley}, {Shapiro}, {Tacconi}, {Buschkamp},
  {Cimatti}, {Daddi}, {Davies}, {Eisenhauer}, {Erb}, {Genel}, {Gerhard},
  {Hicks}, {Lutz}, {Naab}, {Ott}, {Rabien}, {Renzini}, {Steidel}, {Sternberg},
  \& {Lilly}}]{genzel08}
{Genzel}, R., {Burkert}, A., {Bouch{\'e}}, N., {et~al.} 2008, \apj, 687, 59

\bibitem[{{Glazebrook}(2013)}]{glazebrook13}
{Glazebrook}, K. 2013, \pasa, 30, e056

\bibitem[{Grand {et~al.}(2020)Grand, Kawata, Belokurov, Deason, Fattahi,
  Fragkoudi, Gómez, Marinacci, \& Pakmor}]{grand20}
Grand, R. J.~J., Kawata, D., Belokurov, V., {et~al.} 2020, Monthly Notices of
  the Royal Astronomical Society, 497, 1603.
\newblock \url{https://doi.org/10.1093/mnras/staa2057}

\bibitem[{{Guo} {et~al.}(2015){Guo}, {Ferguson}, {Bell}, {Koo}, {Conselice},
  {Giavalisco}, {Kassin}, {Lu}, {Lucas}, {Mandelker}, {McIntosh}, {Primack},
  {Ravindranath}, {Barro}, {Ceverino}, {Dekel}, {Faber}, {Fang}, {Koekemoer},
  {Noeske}, {Rafelski}, \& {Straughn}}]{guo15}
{Guo}, Y., {Ferguson}, H.~C., {Bell}, E.~F., {et~al.} 2015, \apj, 800, 39

\bibitem[{{Hamilton-Campos} {et~al.}(2023{\natexlab{a}}){Hamilton-Campos},
  {Simons}, {Peeples}, {Snyder}, \& {Heckman}}]{hamilton23}
{Hamilton-Campos}, K.~A., {Simons}, R.~C., {Peeples}, M.~S., {Snyder}, G.~F.,
  \& {Heckman}, T.~M. 2023{\natexlab{a}}, \apj, 956, 147

\bibitem[{{Hamilton-Campos} {et~al.}(2023{\natexlab{b}}){Hamilton-Campos},
  {Simons}, {Peeples}, {Snyder}, \& {Heckman}}]{hamiltoncampos23}
---. 2023{\natexlab{b}}, \apj, 956, 147

\bibitem[{{Huang} {et~al.}(2025){Huang}, {Baltasar}, {Ratier-Werbin},
  {Storfer}, {Sheu}, {Agarwal}, {Tamargo-Arizmendi}, {Schlegel}, {Aguilar},
  {Ahlen}, {Aldering}, {Banka}, {BenZvi}, {Bianchi}, {Bolton}, {Brooks},
  {Cikota}, {Claybaugh}, {de la Macorra}, {Dey}, {Doel}, {Edelstein}, {Filipp},
  {Forero-Romero}, {Gaztanaga}, {Gontcho}, {Gu}, {Gutierrez}, {Honscheid},
  {Jullo}, {Juneau}, {Kehoe}, {Kirkby}, {Kisner}, {Kremin}, {Kwon}, {Lambert},
  {Landriau}, {Lang}, {Le Guillou}, {Liu}, {Meisner}, {Miquel}, {Moustakas},
  {Myers}, {Perlmutter}, {Perez-Rafols}, {Prada}, {Rossi}, {Rubin}, {Sanchez},
  {Schubnell}, {Shu}, {Silver}, {Sprayberry}, {Suzuki}, {Tarle}, {Weaver}, \&
  {Zou}}]{huang25}
{Huang}, X., {Baltasar}, S., {Ratier-Werbin}, N., {et~al.} 2025, arXiv
  e-prints, arXiv:2502.03455

\bibitem[{{Huertas-Company} {et~al.}(2015){Huertas-Company},
  {P{\'e}rez-Gonz{\'a}lez}, {Mei}, {Shankar}, {Bernardi}, {Daddi}, {Barro},
  {Cabrera-Vives}, {Cattaneo}, {Dimauro}, \& {Gravet}}]{huertas15}
{Huertas-Company}, M., {P{\'e}rez-Gonz{\'a}lez}, P.~G., {Mei}, S., {et~al.}
  2015, \apj, 809, 95

\bibitem[{{Hung} {et~al.}(2019){Hung}, {Hayward}, {Yuan}, {Boylan-Kolchin},
  {Faucher-Gigu{\`e}re}, {Hopkins}, {Kere{\v{s}}}, {Murray}, \&
  {Wetzel}}]{hung19}
{Hung}, C.-L., {Hayward}, C.~C., {Yuan}, T., {et~al.} 2019, \mnras, 482, 5125

\bibitem[{Inoue \& Saitoh(2014)}]{inoue14}
Inoue, S., \& Saitoh, T.~R. 2014, Monthly Notices of the Royal Astronomical
  Society, 441, 243.
\newblock \url{https://doi.org/10.1093/mnras/stu544}

\bibitem[{{Jacobs} {et~al.}(2023){Jacobs}, {Glazebrook}, {Calabr{\`o}}, {Treu},
  {Nannayakkara}, {Jones}, {Merlin}, {Abraham}, {Stevens}, {Vulcani}, {Yang},
  {Bonchi}, {Boyett}, {Brada{\v{c}}}, {Castellano}, {Fontana}, {Marchesini},
  {Malkan}, {Mason}, {Morishita}, {Paris}, {Santini}, {Trenti}, \&
  {Wang}}]{jacobs23}
{Jacobs}, C., {Glazebrook}, K., {Calabr{\`o}}, A., {et~al.} 2023, \apjl, 948,
  L13

\bibitem[{{Johnson} {et~al.}(2018){Johnson}, {Harrison}, {Swinbank}, {Tiley},
  {Stott}, {Bower}, {Smail}, {Bunker}, {Sobral}, {Turner}, {Best}, {Bureau},
  {Cirasuolo}, {Jarvis}, {Magdis}, {Sharples}, {Bland-Hawthorn}, {Catinella},
  {Cortese}, {Croom}, {Federrath}, {Glazebrook}, {Sweet}, {Bryant}, {Goodwin},
  {Konstantopoulos}, {Lawrence}, {Medling}, {Owers}, \& {Richards}}]{johnson18}
{Johnson}, H.~L., {Harrison}, C.~M., {Swinbank}, A.~M., {et~al.} 2018, \mnras,
  474, 5076

\bibitem[{{Jones} {et~al.}(2010){Jones}, {Swinbank}, {Ellis}, {Richard}, \&
  {Stark}}]{jones10a}
{Jones}, T.~A., {Swinbank}, A.~M., {Ellis}, R.~S., {Richard}, J., \& {Stark},
  D.~P. 2010, \mnras, 404, 1247

\bibitem[{{Kassin} {et~al.}(2012){Kassin}, {Weiner}, {Faber}, {Gardner},
  {Willmer}, {Coil}, {Cooper}, {Devriendt}, {Dutton}, {Guhathakurta}, {Koo},
  {Metevier}, {Noeske}, \& {Primack}}]{kassin12}
{Kassin}, S.~A., {Weiner}, B.~J., {Faber}, S.~M., {et~al.} 2012, \apj, 758, 106

\bibitem[{{Keerthi Vasan G.} {et~al.}(2024){Keerthi Vasan G.}, {Jones},
  {Shajib}, {Rhoades}, {Chen}, {Sanders}, {Stark}, {Ellis}, {Leethochawalit},
  {Kacprzak}, {Barone}, {Glazebrook}, {Tran}, {Skobe}, {Mortensen}, \&
  {Barisic}}]{vasan24}
{Keerthi Vasan G.}, C., {Jones}, T., {Shajib}, A.~J., {et~al.} 2024, arXiv
  e-prints, arXiv:2402.00942

\bibitem[{{Kuhn} {et~al.}(2024){Kuhn}, {Guo}, {Martin}, {Bayless}, {Gates}, \&
  {Puleo}}]{kuhn24}
{Kuhn}, V., {Guo}, Y., {Martin}, A., {et~al.} 2024, \apjl, 968, L15

\bibitem[{Lacey(1984)}]{lacey84}
Lacey, C.~G. 1984, Monthly Notices of the Royal Astronomical Society, 208, 687.
\newblock \url{https://doi.org/10.1093/mnras/208.4.687}

\bibitem[{{Mainali} {et~al.}(2023){Mainali}, {Stark}, {Jones}, {Ellis},
  {Hezaveh}, \& {Rigby}}]{mainali23}
{Mainali}, R., {Stark}, D.~P., {Jones}, T., {et~al.} 2023, \mnras, 520, 4037

\bibitem[{{Martinsson} {et~al.}(2013){Martinsson}, {Verheijen}, {Westfall},
  {Bershady}, {Schechtman-Rook}, {Andersen}, \& {Swaters}}]{martinsson13}
{Martinsson}, T. P.~K., {Verheijen}, M. A.~W., {Westfall}, K.~B., {et~al.}
  2013, \aap, 557, A130

\bibitem[{{Morrissey} {et~al.}(2018){Morrissey}, {Matuszewski}, {Martin},
  {Neill}, {Epps}, {Fucik}, {Weber}, {Darvish}, {Adkins}, {Allen}, {Bartos},
  {Belicki}, {Cabak}, {Callahan}, {Cowley}, {Crabill}, {Deich}, {Delecroix},
  {Doppman}, {Hilyard}, {James}, {Kaye}, {Kokorowski}, {Kwok}, {Lanclos},
  {Milner}, {Moore}, {O'Sullivan}, {Parihar}, {Park}, {Phillips}, {Rizzi},
  {Rockosi}, {Rodriguez}, {Salaun}, {Seaman}, {Sheikh}, {Weiss}, \&
  {Zarzaca}}]{morrissey18}
{Morrissey}, P., {Matuszewski}, M., {Martin}, D.~C., {et~al.} 2018, \apj, 864,
  93

\bibitem[{{Mortensen} {et~al.}(2021){Mortensen}, {Keerthi Vasan}, {Jones},
  {Faucher-Gigu{\`e}re}, {Sanders}, {Ellis}, {Leethochawalit}, \&
  {Stark}}]{mortensen21}
{Mortensen}, K., {Keerthi Vasan}, G.~C., {Jones}, T., {et~al.} 2021, \apj, 914,
  92

\bibitem[{{Newman} {et~al.}(2018){Newman}, {Belli}, {Ellis}, \&
  {Patel}}]{newman18}
{Newman}, A.~B., {Belli}, S., {Ellis}, R.~S., \& {Patel}, S.~G. 2018, \apj,
  862, 126

\bibitem[{{Newman} {et~al.}(2010){Newman}, {Ellis}, {Treu}, \&
  {Bundy}}]{newman10}
{Newman}, A.~B., {Ellis}, R.~S., {Treu}, T., \& {Bundy}, K. 2010, \apjl, 717,
  L103

\bibitem[{{Pessa} {et~al.}(2023){Pessa}, {Schinnerer}, {Sanchez-Blazquez},
  {Belfiore}, {Groves}, {Emsellem}, {Neumann}, {Leroy}, {Bigiel}, {Chevance},
  {Dale}, {Glover}, {Grasha}, {Klessen}, {Kreckel}, {Kruijssen}, {Pinna},
  {Querejeta}, {Rosolowsky}, \& {Williams}}]{pessa23}
{Pessa}, I., {Schinnerer}, E., {Sanchez-Blazquez}, P., {et~al.} 2023, \aap,
  673, A147

\bibitem[{{Rizzo} {et~al.}(2023){Rizzo}, {Roman-Oliveira}, {Fraternali},
  {Frickmann}, {Valentino}, {Brammer}, {Zanella}, {Kokorev}, {Popping},
  {Whitaker}, {Kohandel}, {Magdis}, {Di Mascolo}, {Ikeda}, {Jin}, \&
  {Toft}}]{rizzo23}
{Rizzo}, F., {Roman-Oliveira}, F., {Fraternali}, F., {et~al.} 2023, \aap, 679,
  A129

\bibitem[{{Rizzo} {et~al.}(2024){Rizzo}, {Bacchini}, {Kohandel}, {Di Mascolo},
  {Fraternali}, {Roman-Oliveira}, {Zanella}, {Popping}, {Valentino}, {Magdis},
  \& {Whitaker}}]{rizzo24}
{Rizzo}, F., {Bacchini}, C., {Kohandel}, M., {et~al.} 2024, \aap, 689, A273

\bibitem[{{Scoville} {et~al.}(2016){Scoville}, {Sheth}, {Aussel}, {Vanden
  Bout}, {Capak}, {Bongiorno}, {Casey}, {Murchikova}, {Koda},
  {{\'A}lvarez-M{\'a}rquez}, {Lee}, {Laigle}, {McCracken}, {Ilbert}, {Pope},
  {Sanders}, {Chu}, {Toft}, {Ivison}, \& {Manohar}}]{scoville16}
{Scoville}, N., {Sheth}, K., {Aussel}, H., {et~al.} 2016, \apj, 820, 83

\bibitem[{{Shapley}(2011)}]{shapley11}
{Shapley}, A.~E. 2011, \araa, 49, 525

\bibitem[{{Shapley} {et~al.}(2003){Shapley}, {Steidel}, {Pettini}, \&
  {Adelberger}}]{shapley03}
{Shapley}, A.~E., {Steidel}, C.~C., {Pettini}, M., \& {Adelberger}, K.~L. 2003,
  \apj, 588, 65

\bibitem[{{Simons} {et~al.}(2017){Simons}, {Kassin}, {Weiner}, {Faber},
  {Trump}, {Heckman}, {Koo}, {Pacifici}, {Primack}, {Snyder}, \& {de la
  Vega}}]{simons17}
{Simons}, R.~C., {Kassin}, S.~A., {Weiner}, B.~J., {et~al.} 2017, \apj, 843, 46

\bibitem[{{Soto} {et~al.}(2012){Soto}, {Martin}, {Prescott}, \&
  {Armus}}]{soto2012}
{Soto}, K.~T., {Martin}, C.~L., {Prescott}, M.~K.~M., \& {Armus}, L. 2012,
  \apj, 757, 86

\bibitem[{{Stark} {et~al.}(2008){Stark}, {Swinbank}, {Ellis}, {Dye}, {Smail},
  \& {Richard}}]{stark08}
{Stark}, D.~P., {Swinbank}, A.~M., {Ellis}, R.~S., {et~al.} 2008, \nat, 455,
  775

\bibitem[{{Swinbank} {et~al.}(2011){Swinbank}, {Papadopoulos}, {Cox}, {Krips},
  {Ivison}, {Smail}, {Thomson}, {Neri}, {Richard}, \& {Ebeling}}]{swinbank11}
{Swinbank}, A.~M., {Papadopoulos}, P.~P., {Cox}, P., {et~al.} 2011, \apj, 742,
  11

\bibitem[{{Tacconi} {et~al.}(2020){Tacconi}, {Genzel}, \&
  {Sternberg}}]{tacconi20}
{Tacconi}, L.~J., {Genzel}, R., \& {Sternberg}, A. 2020, \araa, 58, 157

\bibitem[{{Tacconi} {et~al.}(2013){Tacconi}, {Neri}, {Genzel}, {Combes},
  {Bolatto}, {Cooper}, {Wuyts}, {Bournaud}, {Burkert}, {Comerford}, {Cox},
  {Davis}, {F{\"o}rster Schreiber}, {Garc{\'\i}a-Burillo}, {Gracia-Carpio},
  {Lutz}, {Naab}, {Newman}, {Omont}, {Saintonge}, {Shapiro Griffin}, {Shapley},
  {Sternberg}, \& {Weiner}}]{tacconi13}
{Tacconi}, L.~J., {Neri}, R., {Genzel}, R., {et~al.} 2013, \apj, 768, 74

\bibitem[{{Toft} {et~al.}(2017){Toft}, {Zabl}, {Richard}, {Gallazzi},
  {Zibetti}, {Prescott}, {Grillo}, {Man}, {Lee}, {G{\'o}mez-Guijarro},
  {Stockmann}, {Magdis}, \& {Steinhardt}}]{toft17}
{Toft}, S., {Zabl}, J., {Richard}, J., {et~al.} 2017, \nat, 546, 510

\bibitem[{{Tsukui} {et~al.}(2024){Tsukui}, {Wisnioski}, {Bland-Hawthorn}, \&
  {Freeman}}]{tsukui24}
{Tsukui}, T., {Wisnioski}, E., {Bland-Hawthorn}, J., \& {Freeman}, K. 2024,
  arXiv e-prints, arXiv:2409.15909

\bibitem[{{Turner} {et~al.}(2017){Turner}, {Cirasuolo}, {Harrison}, {McLure},
  {Dunlop}, {Swinbank}, {Johnson}, {Sobral}, {Matthee}, \&
  {Sharples}}]{turner17}
{Turner}, O.~J., {Cirasuolo}, M., {Harrison}, C.~M., {et~al.} 2017, \mnras,
  471, 1280

\bibitem[{{{\"U}bler} {et~al.}(2018){{\"U}bler}, {Genzel}, {Tacconi},
  {F{\"o}rster Schreiber}, {Neri}, {Contursi}, {Belli}, {Nelson}, {Lang},
  {Shimizu}, {Davies}, {Herrera-Camus}, {Lutz}, {Plewa}, {Price}, {Schuster},
  {Sternberg}, {Tadaki}, {Wisnioski}, \& {Wuyts}}]{ubler18}
{{\"U}bler}, H., {Genzel}, R., {Tacconi}, L.~J., {et~al.} 2018, \apjl, 854, L24

\bibitem[{{{\"U}bler} {et~al.}(2019){{\"U}bler}, {Genzel}, {Wisnioski},
  {F{\"o}rster Schreiber}, {Shimizu}, {Price}, {Tacconi}, {Belli}, {Wilman},
  {Fossati}, {Mendel}, {Davies}, {Beifiori}, {Bender}, {Brammer}, {Burkert},
  {Chan}, {Davies}, {Fabricius}, {Galametz}, {Herrera-Camus}, {Lang}, {Lutz},
  {Momcheva}, {Naab}, {Nelson}, {Saglia}, {Tadaki}, {van Dokkum}, \&
  {Wuyts}}]{ubler19}
{{\"U}bler}, H., {Genzel}, R., {Wisnioski}, E., {et~al.} 2019, \apj, 880, 48

\bibitem[{{{\"U}bler} {et~al.}(2024){{\"U}bler}, {F{\"o}rster Schreiber}, {van
  der Wel}, {Bezanson}, {Price}, {D'Eugenio}, {Wisnioski}, {Genzel}, {Tacconi},
  {Wuyts}, {Naab}, {Lutz}, {Straatman}, {Shimizu}, {Davies}, {Liu}, \&
  {Mendel}}]{ubler24}
{{\"U}bler}, H., {F{\"o}rster Schreiber}, N.~M., {van der Wel}, A., {et~al.}
  2024, \mnras, 527, 9206

\bibitem[{van Donkelaar {et~al.}(2022)van Donkelaar, Agertz, \&
  Renaud}]{vandonkelaar22}
van Donkelaar, F., Agertz, O., \& Renaud, F. 2022, Monthly Notices of the Royal
  Astronomical Society, 512, 3806.
\newblock \url{https://doi.org/10.1093/mnras/stac692}

\bibitem[{{Wisnioski} {et~al.}(2015){Wisnioski}, {F{\"o}rster Schreiber},
  {Wuyts}, {Wuyts}, {Bandara}, {Wilman}, {Genzel}, {Bender}, {Davies},
  {Fossati}, {Lang}, {Mendel}, {Beifiori}, {Brammer}, {Chan}, {Fabricius},
  {Fudamoto}, {Kulkarni}, {Kurk}, {Lutz}, {Nelson}, {Momcheva}, {Rosario},
  {Saglia}, {Seitz}, {Tacconi}, \& {van Dokkum}}]{wisnioski15}
{Wisnioski}, E., {F{\"o}rster Schreiber}, N.~M., {Wuyts}, S., {et~al.} 2015,
  \apj, 799, 209

\bibitem[{{Wisnioski} {et~al.}(2019){Wisnioski}, {F{\"o}rster Schreiber},
  {Fossati}, {Mendel}, {Wilman}, {Genzel}, {Bender}, {Wuyts}, {Davies},
  {{\"U}bler}, {Bandara}, {Beifiori}, {Belli}, {Brammer}, {Chan}, {Davies},
  {Fabricius}, {Galametz}, {Lang}, {Lutz}, {Nelson}, {Momcheva}, {Price},
  {Rosario}, {Saglia}, {Seitz}, {Shimizu}, {Tacconi}, {Tadaki}, {van Dokkum},
  \& {Wuyts}}]{wisnioski19}
{Wisnioski}, E., {F{\"o}rster Schreiber}, N.~M., {Fossati}, M., {et~al.} 2019,
  \apj, 886, 124

\bibitem[{{Wuyts} {et~al.}(2012){Wuyts}, {F{\"o}rster Schreiber}, {Genzel},
  {Guo}, {Barro}, {Bell}, {Dekel}, {Faber}, {Ferguson}, {Giavalisco}, {Grogin},
  {Hathi}, {Huang}, {Kocevski}, {Koekemoer}, {Koo}, {Lotz}, {Lutz}, {McGrath},
  {Newman}, {Rosario}, {Saintonge}, {Tacconi}, {Weiner}, \& {van der
  Wel}}]{wuyts12}
{Wuyts}, S., {F{\"o}rster Schreiber}, N.~M., {Genzel}, R., {et~al.} 2012, \apj,
  753, 114

\bibitem[{{Yoachim} \& {Dalcanton}(2006)}]{yoachim06}
{Yoachim}, P., \& {Dalcanton}, J.~J. 2006, \aj, 131, 226

\bibitem[{Yu {et~al.}(2023)Yu, Bullock, Gurvich, Hafen, Stern, Boylan-Kolchin,
  Faucher-Giguère, Wetzel, Hopkins, \& Moreno}]{yu23}
Yu, S., Bullock, J.~S., Gurvich, A.~B., {et~al.} 2023, Monthly Notices of the
  Royal Astronomical Society, 523, 6220.
\newblock \url{https://doi.org/10.1093/mnras/stad1806}

\end{thebibliography}

\end{document}